\documentclass[reprint,showpacs,preprintnumbers,amsmath,amssymb,floatfix,prc]{revtex4-1}
\usepackage{graphicx}
\usepackage{dcolumn}
\usepackage{bm}

\def\beq{\begin{equation}}
\def\eeq{\end{equation}}
\def\be{\begin{eqnarray}}
\def\ee{\end{eqnarray}}

\newcommand{\lsim}{
 \mathrel{\setbox0=\hbox{$<$}\raise0.6ex\copy0\kern-\wd0
 \lower0.65ex\hbox{$\sim$}}}
 
\newcommand{\gsim}{
 \mathrel{\setbox0=\hbox{$>$}\raise0.6ex\copy0\kern-\wd0
 \lower0.65ex\hbox{$\sim$}}}

\begin{document}
\title{Coherent photoproduction of $\psi$ and $\Upsilon$
mesons in ultraperipheral pPb and PbPb collisions at the CERN LHC}
\author{Adeola Adeluyi}
\affiliation{Department of Physics \& Astronomy,
Texas A\&M University-Commerce, Commerce, TX 75428, USA}
\author{Trang Nguyen}
\affiliation{Center for Nuclear Research, Department of Physics \\
Kent State University, Kent, OH 44242, USA}
\date{\today}
\begin{abstract}
Exclusive photoproduction of vector mesons in the perturbative 
two-gluon exchange formalism depends significantly on nucleon and 
nuclear gluon distributions. In the present study 
we calculate total cross sections and rapidity distributions of 
$J/\psi(1s)$, $\psi(2s)$, $\Upsilon(1s)$, $\Upsilon(2s)$, and 
$\Upsilon(3s)$ in ultraperipheral proton-lead (pPb) and 
lead-lead (PbPb) collisions at the CERN Large Hadron Collider (LHC)
at $\sqrt{s_{_{NN}}}=5$ TeV and $\sqrt{s_{_{NN}}}=2.76$ TeV respectively. 
Effects of gluon shadowing are investigated and potentials for
constraining nuclear gluon modifications are discussed.
\end{abstract}
\maketitle
\vspace{1cm}
%
%
Photoproduction of heavy quarkonia in ultraperipheral collisions 
can help elucidate several aspects of strong interaction dynamics at 
high energies. It is an important part of current experimental efforts 
at the CERN Large Hadron Collider (LHC). All four of the large LHC experiments, 
ALICE, ATLAS, CMS and LHCb, have the capability to measure heavy 
quarkonia and first results on coherent photoproduction of $J/\psi(1s)$ in 
ultraperipheral PbPb collisions at $\sqrt{s_{_{NN}}}=2.76$ TeV) have 
recently been presented by the ALICE Collaboration \cite{Abelev:2012ba}.
On the theory front it has been extensively studied at various 
energies and for different collision systems (see for instance 
\cite{Lappi:2013am,Cisek:2012yt,Rebyakova:2011vf,Ahmadov:2012dn,Adeluyi:2012ph,
Adeluyi:2012ds,Adeluyi:2011rt,Goncalves:2011vf,AyalaFilho:2008zr, 
Goncalves:2007sa,Ivanov:2007ms,Goncalves:2005ge,Goncalves:2005sn, 
Klein:2003vd,Frankfurt:2003qy,Frankfurt:2001db,Klein:1999qj}). 
Diverse approaches ranging from models based on perturbative Quantum 
Chromodynamics (pQCD) to Color Dipole models and $k_T$ factorization 
have been used to study coherent and incoherent photoproduction of
heavy mesons. The characteristics of some of the currently employed
models is discussed in \cite{Abelev:2012ba}.

In previous works \cite{Adeluyi:2012ph,Adeluyi:2012ds,Adeluyi:2011rt} 
we have considered coherent photoproduction of $J/\psi(1s)$ and 
$\Upsilon(1s)$ at various energies in pPb and PbPb collisions at the
LHC. The calculations were carried out in the framework of
perturbative two-gluon exchange formalism with different nuclear 
gluon distributions and a detailed exposition can be found in 
\cite{Adeluyi:2012ph} (see also \cite{Ryskin,Ryskin:1995hz}). In the 
current article we extend these previous studies by considering, in 
addition to $J/\psi(1s)$ and $\Upsilon(1s)$,  
the exclusive photoproduction of $\psi(2s)$, $\Upsilon(2s)$, and 
$\Upsilon(3s)$ in ultraperipheral pPb (at $\sqrt{s_{_{NN}}}=5$ TeV) and PbPb 
(at $\sqrt{s_{_{NN}}}=2.76$ TeV) collisions at the LHC. For brevity we will  
refer to $J/\psi(1s)$ and $\psi(2s)$ collectively as $\psi$ mesons, and
$\Upsilon(1s)$, $\Upsilon(2s)$, and $\Upsilon(3s)$ as $\Upsilon$ mesons.
The calculation framework remains essentially unchanged;
the only additional assumption is that the multiplicative correction 
factor $\zeta_V$ introduced in \cite{Adeluyi:2012ph} and determined
for $J/\psi(1s)$ is applicable to $\psi(2s)$. The requisite input
masses and widths of these mesons are taken from \cite{Beringer:1900zz} 
and are shown in Table~\ref{mesmasgam}. As in previous studies 
we use gluon distributions from MSTW08 \cite{Martin:2009iq}, EPS08 
\cite{Eskola:2008ca}, EPS09 \cite{Eskola:2009uj}, and 
HKN07 \cite{Hirai:2007sx}.  The characteristics of these
distributions, especially the disparities in
the strength of the nuclear modifications of their gluon content, have
been treated in detail in \cite{Adeluyi:2012ph}. Summarily, in terms
of gluon shadowing strength, the progression is from zero effects 
(MSTW08) to weak effects (HKN07), moderate effects (EPS09), and strong 
effects (EPS08). From the collision energies involved, the current
work can be regarded as a natural extension of the study reported 
in \cite{Adeluyi:2012ds}, and some similarities in features are thus
to be expected.
\begin{table}[!htb]
\caption{\label{mesmasgam} Masses $M_V$ and leptonic decay 
widths $\Gamma_{ee}$ of $\psi$ and $\Upsilon$ mesons. 
Data taken from \cite{Beringer:1900zz}.}
\begin{tabular}[c]{|l | c| c| c| c|}
\hline
\hline
 meson               & $M_V$ (GeV)   & $\Gamma_{ee}$ (KeV)\\
\hline
\hline
$J/\psi(1s)$         & 3.096916      & 5.55 \\ 
$\psi(2s)$           & 3.686108      & 2.33 \\
\hline
\hline
$\Upsilon(1s)$       & 9.4603        & 1.34 \\ 
$\Upsilon(2s)$       & 10.02326      & 0.612  \\
$\Upsilon(3s)$       & 10.3352       & 0.443  \\
\hline
\hline
\end{tabular} 
\end{table}

\begin{figure}[!h]
\includegraphics[width=\columnwidth]{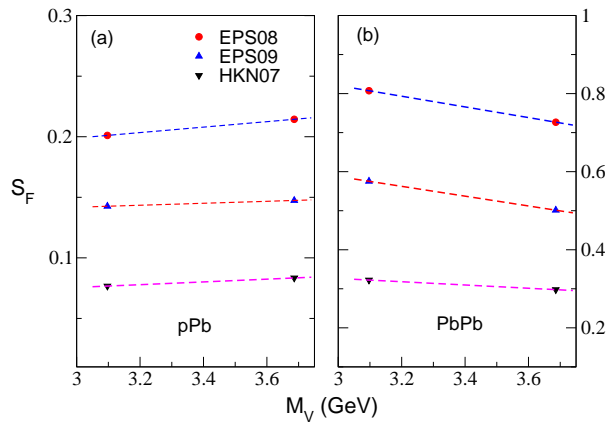}
\caption[...]{\label{shdratpsi} (Color online) Shadowing factor $S_F$
for photoproduction of $J/\psi(1s)$ and $\psi(2s)$ in (a) pPb and (b) 
PbPb collisions at the LHC. Dashed lines are for visual clarity.}
\end{figure}

Let us now present the cross sections and rapidity
distributions of $J/\psi(1s)$ and $\psi(2s)$ in ultraperipheral 
pPb and PbPb collisions. Upper panel of Table~\ref{tpsipPbPbPb} shows 
the total cross sections in $\mu$b for elastic photoproduction of 
these mesons in ultraperipheral pPb collisions while the lower panel displays 
the corresponding cross sections in mb for PbPb collisions. 
\begin{table}[!htb]
\caption{\label{tpsipPbPbPb} Total cross sections for elastic 
photoproduction of $J/\psi(1s)$ and $\psi(2s)$ in ultraperipheral pPb 
(in $\mu$b; upper panel) and 
PbPb (in mb; lower panel) collisions at the LHC.}
\begin{tabular}[c]{|l | c| c| c| c|}
\hline
\hline
 Meson          & MSTW08  & EPS08  & EPS09   & HKN07 \\
\hline
\hline
$J/\psi(1s)$    & 86.5    & 69.1   & 74.2    & 79.9 \\ 
$\psi(2s)$      & 18.6    & 14.6   & 15.8    & 17.0 \\
\hline
\hline
$J/\psi(1s)$    & 34.4    & 6.6    & 14.6    & 23.3 \\ 
$\psi(2s)$      & 7.1     & 1.9    & 3.5     & 5.0 \\
\hline
\hline
\end{tabular} 
\end{table}
The cross sections for each meson reflect the trend in gluon shadowing
strength: those from MSTW08 are the largest while
the EPS08 cross sections are the smallest. A quantitative 
measure of the overall effect of gluon shadowing on total 
photoproduction cross section can be obtained by defining, 
for each meson $V$ and nuclear gluon distribution NGD, a shadowing factor, 
$S_F$, given by 
\begin{equation}
S_F = \frac{\sigma^{MSTW08}_{_V}-\sigma^{NGD}_{_V}}{\sigma^{MSTW08}_{_V}}
\label{shadfact}
\end{equation}
As defined, $S_F$ scales linearly with severity of shadowing,
i.e. larger values of $S_F$ translate to larger shadowing effects.
When multiplied by $100$ the resulting product gives the percentage 
by which the no-shadowing MSTW08 cross section is reduced by the
shadowing in the specified nuclear gluon distribution.
$S_F$ for both mesons and all three nuclear gluon distributions are shown in 
Fig.~\ref{shdratpsi} for pPb collisions (left panel) and PbPb
collisions (right panel) respectively.     

Let us first consider pPb collisions (left panel). Shadowing effects 
are most pronounced for EPS08 and least pronounced for HKN07 and
almost the same for both mesons. As can be seen from Fig.~\ref{shdratpsi} 
there is approximately a $20\%$ reduction of the MSTW08 cross section
by the shadowing effects in EPS08, $14\%$ by EPS09 and $7\%$ by HKN07 
respectively. 

Shadowing effects are significantly stronger in PbPb collisions as 
evidenced by the larger values of $S_F$ and follow the usual trend: 
they are largest for EPS08 and smallest for HKN07. 
Also $J/\psi(1s)$ exhibit greater sensitivity to shadowing effects 
than $\psi(2s)$, especially for EPS08. For $J/\psi(1s)$ the reduction 
ranges from about $32\%$ (HKN07) to about $80\%$ (EPS08) while for 
$\psi(2s)$ it ranges from $30\%$ to $72\%$.

In Table~\ref{rpsipPbPbPb} we show the ratio 
$\sigma^{\psi(2s)}/\sigma^{J/\psi(1s)}$ from all four gluon
distributions for pPb and PbPb collisions. It is $\approx 0.21$ for 
pPb collisions and between $0.21$ and $0.29$ for PbPb collisions, in 
line with the trend seen in $S_F$.
\begin{table}[!htb]
\caption{\label{rpsipPbPbPb} Ratio of cross sections for pPb 
(upper row) and PbPb (lower row) collisions.}
\begin{tabular}[c]{|l | c| c| c| c|}
\hline
\hline
Ratio           & MSTW08  & EPS08  & EPS09   & HKN07 \\
\hline
\hline
$\sigma^{\psi(2s)}/\sigma^{J/\psi(1s)}$    & 0.215    & 0.211   & 0.213    & 0.213 \\ 
\hline
\hline
$\sigma^{\psi(2s)}/\sigma^{J/\psi(1s)}$    & 0.206    & 0.288    & 0.240    & 0.215 \\ 
\hline
\hline
\end{tabular} 
\end{table}

\begin{figure}[!htb]
\includegraphics[width=\columnwidth]{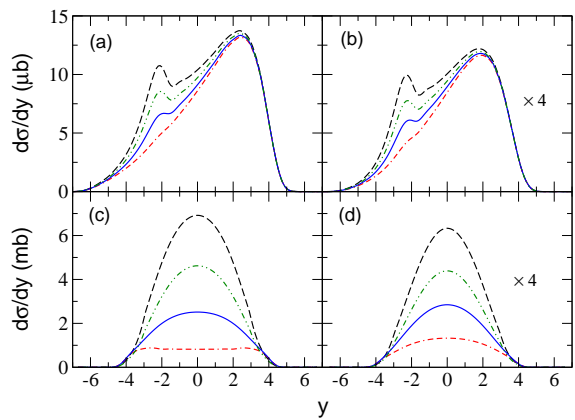}
\caption[...]{\label{psirap} (Color online) Rapidity distributions
of exclusive photoproduction of $J/\psi(1s)$ ((a) and (c)) and 
$\psi(2s)$ ((b) and (d)) in pPb (upper panels) and PbPb (lower panels)
collisions at the LHC. Dashed (MSTW08), dash-double-dotted (HKN07), solid 
(EPS09), and dash-dotted (EPS08) lines correspond to rapidity 
distributions with no shadowing, weak shadowing, moderate shadowing, 
and strong shadowing respectively. The $\psi(2s)$ 
distributions have been magnified for clarity.}
\end{figure}

We now consider rapidity distributions. 
Fig.~\ref{psirap} shows the rapidity distributions for $J/\psi(1s)$
and $\psi(2s)$ in pPb (upper panel) and PbPb (lower panel)   
collisions. For pPb collisions the distributions shown 
are the sum of the  $\gamma$p and  $\gamma$Pb contributions
and are manifestly asymmetric in line with the convention
adopted in \cite{Adeluyi:2012ph}.
The rapidity distributions exhibit clearly the influence of 
gluon shadowing in the interval $-4 \lesssim y \lesssim 2$, and 
especially in the narrow rapidity window 
$-3 \lesssim y \lesssim -1$ where the differences reflect the 
relative strength of gluon shadowing in the respective nuclear 
parton distribution. Also the shapes are similar for both mesons.
Thus it seems feasible that a consideration of 
especially $J/\psi(1s)$ production in pPb collisions in 
this rapidity interval offers some potential in constraining gluon shadowing. 

The rapidity distributions in PbPb collisions 
are symmetric about midrapidity and also similar in structure 
except for EPS08. The influence of the significantly stronger nuclear effects  
in PbPb collisions is demonstrated by the remarkable distinctions
in the rapidity distributions predicted by the different gluon
distributions over an appreciable range of rapidity.
Shadowing is the relevant nuclear effect in the rapidity interval 
$-3 < y < 3$ and the rapidity distributions mimic 
the shadowing strength of the various distributions. 
In particular the rapidity window $-2 < y < 2$ manifestly 
depicts the significant distinctions between the various gluon 
distributions. This interval is thus suitable for probing gluon shadowing.
The influence of antishadowing is manifested in the intervals $-4.5 < y
< -3.5$ and $3.5 < y <4.5$ but the effect is relatively slight.      

Let us now turn to $\Upsilon$ mesons in ultraperipheral pPb 
and PbPb collisions. The ensuing treatment parallels closely that of 
the $\psi$ mesons in many respects due to the same 
underlying production mechanism.

\begin{figure}[!htb]
\includegraphics[width=\columnwidth]{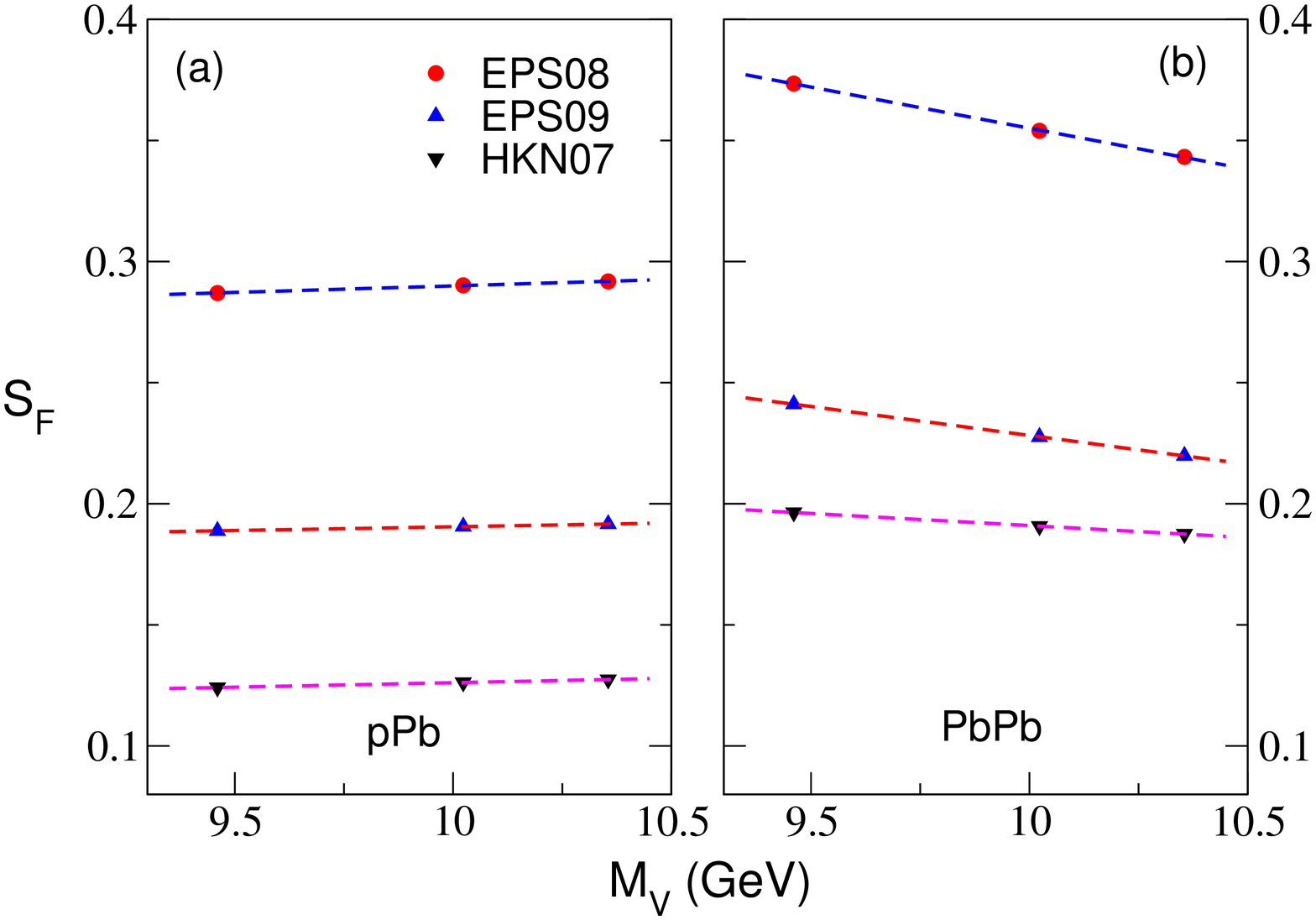}
\caption[...]{\label{shdratups} (Color online) Shadowing factor $S_F$
for photoproduction of $\Upsilon(1s)$, $\Upsilon(2s)$, and
$\Upsilon( 3s)$ in (a) pPb and (b) PbPb collisions at the LHC. 
Dashed lines are for visual clarity.}
\end{figure}

\begin{table}[!htb]
\caption{\label{tupspPbPbPb} Total cross sections for elastic 
photoproduction of $\Upsilon$ mesons in ultraperipheral pPb 
(in nb; upper panel) and PbPb (in $\mu$b; lower panel) collisions at the LHC.}
\begin{tabular}[c]{|l | c| c| c| c|}
\hline
\hline
 Meson               & MSTW08   & EPS08    & EPS09    & HKN07 \\
\hline
\hline
$\Upsilon(1s)$       & 291.0   & 207.5   & 236.1   & 254.9 \\ 
$\Upsilon(2s)$       & 94.2    & 66.8    & 76.2    & 82.3 \\
$\Upsilon(3s)$       & 56.1    & 39.7    & 45.4    & 49.0 \\
\hline
\hline
$\Upsilon(1s)$       & 52.1    & 32.7    & 39.5    & 41.9 \\ 
$\Upsilon(2s)$       & 15.9    & 10.3    & 12.3    & 12.9 \\
$\Upsilon(3s)$       & 9.2     & 6.0     & 7.2     & 7.5 \\
\hline
\hline
\end{tabular} 
\end{table}
Upper panel of Table~\ref{tupspPbPbPb} shows the total cross 
sections in nb for elastic photoproduction of $\Upsilon$ mesons in
ultraperipheral pPb collisions while the lower panel displays 
the corresponding cross sections in $\mu$b for PbPb collisions. 
As in the case of $\psi$ mesons the effect of gluon 
shadowing on total cross sections is reflected in the plot of 
$S_F$ as shown in Fig.~\ref{shdratups} for 
pPb collisions (left panel) and PbPb collisions (right panel) 
respectively. 

As is apparent from left panel of Fig.~\ref{shdratups} $S_F$ is approximately 
constant for all three distributions in pPb collisions. Thus the 
no-shadowing (MSTW08) cross sections are reduced by about $29\%$,
$19\%$, and $12\%$ respectively by the shadowing in EPS08, EPS09, and HKN07.
This appreciable magnitude of the effect of shadowing
indicates that $\Upsilon$ photoproduction cross section
in ultraperipheral pPb collisions offers some promising potential in
constraining nuclear gluon shadowing. 

For PbPb collisions $S_F$ is almost constant for HKN07 and decreases
with increasing mass for EPS08 and EPS09. For HKN07 the reduction 
is approximately $19\%$ while for EPS08 and EPS09 it is between.
$37\%$ and $34\%$ and between $24\%$ and $22\%$ respectively.
These reductions are quite significant and thus 
the cross sections for photoproduction of $\Upsilon$ mesons offer 
good constraining ability for gluon shadowing determination.

\begin{table}[!htb]
\caption{\label{rupsipPbPbPb} Ratio of cross sections for pPb 
(upper panel) and PbPb (lower panel) collisions.}
\begin{tabular}[c]{|l | c| c| c| c|}
\hline
\hline
Ratio          & MSTW08  & EPS08  & EPS09   & HKN07 \\
\hline
\hline
$\sigma^{\Upsilon(2s)}/\sigma^{\Upsilon(1s)}$    & 0.324    & 0.322   & 0.323    & 0.323 \\ 
$\sigma^{\Upsilon(3s)}/\sigma^{\Upsilon(1s)}$      & 0.193    & 0.191   & 0.192    & 0.192 \\
\hline
\hline
$\sigma^{\Upsilon(2s)}/\sigma^{\Upsilon(1s)}$    & 0.305    & 0.315    & 0.311    & 0.308 \\ 
$\sigma^{\Upsilon(3s)}/\sigma^{\Upsilon(1s)}$      & 0.177     & 0.183    & 0.182     & 0.179 \\
\hline
\hline
\end{tabular} 
\end{table}
Table~\ref{rupsipPbPbPb} shows the ratios 
$\sigma^{\Upsilon(2s)}/\sigma^{\Upsilon(1s)}$ and 
$\sigma^{\Upsilon(3s)}/\sigma^{\Upsilon(1s)}$ for pPb and PbPb 
collisions. 
Similar ratios in the hadroproduction of $\Upsilon$ in PbPb
collisions have been reported in \cite{Chatrchyan:2011pe,Chatrchyan:2012lxa}. 
The ratios for both collision systems considered are to a good 
approximation independent of shadowing effects.

\begin{figure}[!htb]
\includegraphics[width=\columnwidth]{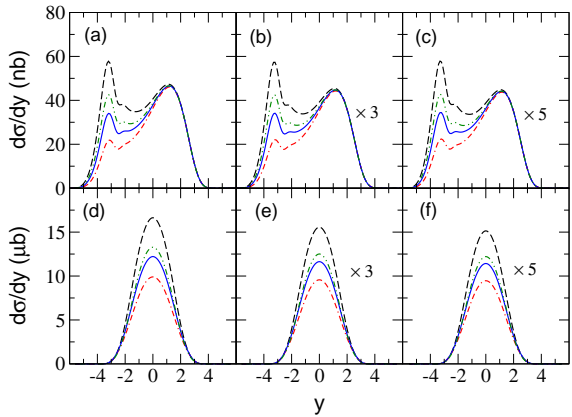}
\caption[...]{\label{upsrap} (Color online) Rapidity distributions 
of exclusive photoproduction of $\Upsilon(1s)$ ((a) and (d)), 
$\Upsilon(2s)$ ((b) and (e)), and  $\Upsilon(3s)$ ((c) and (f)) in 
pPb (upper panels) and PbPb (lower panels) collisions at the LHC.
Dashed (MSTW08), dash-double-dotted (HKN07), solid 
(EPS09), and dash-dotted (EPS08) lines correspond to rapidity 
distributions with no shadowing, weak shadowing, moderate shadowing, 
and strong shadowing respectively. The $\Upsilon(2s)$ and 
$\Upsilon(3s)$ distributions have been magnified for clarity.}
\end{figure}

We now turn to rapidity distributions. 
Fig.~\ref{upsrap} shows the rapidity distributions for 
$\Upsilon$ mesons in pPb (upper panel) and PbPb (lower panel)   
collisions. 
Again for pPb collisions the distributions shown 
are the sum of the  $\gamma$p and $\gamma$Pb contributions. 
 
Let us consider the upper panel. At negative rapidities the distributions  
are relatively more prominent than for $\psi$ production, due to the 
relatively larger influence of the $\gamma$Pb contribution to the
total rapidity distributions. Nuclear effects are clearly discernible
in the interval $-5 \lesssim y \lesssim 0$, and for 
$-4 \lesssim y \lesssim -1$ quite distinctly reflect the effect
of the varying shadowing strength in the gluon distributions used.    
Thus in the rapidity interval $-4 \lesssim y \lesssim -1$ the 
total distributions show good sensitivity to 
gluon shadowing, and therefore afford good potential for constraining 
purposes.

The rapidity distributions for $\Upsilon$ mesons in PbPb collisions 
are symmetric about midrapidity, and are structurally similar.  
The stronger nuclear effects in PbPb collisions again lead
to clear cut differences in the rapidity distributions predicted by the 
different gluon parametrizations considered over a significant
rapidity range.
Shadowing is the relevant nuclear effect in the rapidity interval 
$-2 < y < 2$ and in particular the rapidity window $-1 < y < 1$  
shows clearly the distinctions between the various gluon 
distributions. Therefore this interval is ideal for 
probing gluon shadowing as well as discriminating between the 
different gluon shadowing templates considered in the current study.
The influence of antishadowing is manifested in the intervals $-3 < y
< -2.5$ and $2.5 < y <3$ but the effect is relatively slight.

In conclusion we have considered elastic photoproduction  
of $\psi$ and $\Upsilon$ mesons in ultraperipheral 
pPb and PbPb collisions at LHC.
The production mechanism involves nuclear gluon
distributions and different sets of nuclear 
parton distributions with varying severity of gluon shadowing 
have been utilized. Cross sections, rapidity 
distributions, and cross section ratios for both 
collision systems have been presented.

The significant dependence on gluon distribution implies that
elastic photoproduction of vector mesons in
ultraperipheral collisions could potentially be useful in constraining
modifications such as shadowing in nuclear gluon distributions. 
The cross sections and rapidity distributions for $\psi$ and 
$\Upsilon$ photoproduction in PbPb collisions exhibit significant
sensitivity to gluon shadowing. Thus both offer good potential 
in constraining the shadowing component of nuclear gluon
distributions. In the case of pPb collisions $\Upsilon$, and to a
lesser extent $\psi$ production, also display appreciable sensitivity 
and could thus be of use in constraining purposes. 

%

\end{document}